\documentclass[12pt]{article}
\usepackage[dvips]{graphicx}
\usepackage{pdproc}

  \makeatletter 
  \def\@cite#1{[#1]} 
  \makeatother    
\def\beq{\begin{equation}}
\def\eeq{\end{equation}}
\def\beqa{\begin{eqnarray}}
\def\eeqa{\end{eqnarray}}

\def\lsim{\mathrel{\raise.3ex\hbox{$<$\kern-.75em\lower1ex\hbox{$\sim$}}} }
\def\gsim{\mathrel{\raise.3ex\hbox{$>$\kern-.75em\lower1ex\hbox{$\sim$}}} }

\begin{document}
\thispagestyle{empty}

\begin{flushright}
NCU-HEP-k019  \\
Sep 2004
\end{flushright}

\vspace*{.5in}

\begin{center}
{\bf  \Large Fermion Sector of  Little Higgs $^*$}\\
\vspace*{.5in}
{\bf  Otto C.W. Kong}\\[.05in]
{\it Department of Physics, National Central University, Chung-li, TAIWAN 32054 \\
E-mail: otto@phy.ncu.edu.tw}

\vspace*{.8in}
{\bf Abstract}\\[.2in]
\end{center}
The little Higgs mechanism provides an alternative solution to
the hierarchy problem, arguably fitting better into the phenomenological
hint of the "little hierarchy" which may cause some fine-tuning for
the case of supersymmetry. We discuss an aspect of little Higgs physics
lacking proper attention --- the construction of an interesting and consistent 
chiral fermionic sector and its phenomenological implications. At least for 
the kind of example models to be discussed, the gauge and top sector 
structure of a model largely dictates, through gauge anomaly cancellation 
conditions, a specific chiral fermion spectrum. The spectrum has 
interesting, family non-universal, flavor structure. The implications for 
flavor physics are specially interesting.  We also add a brief comment 
of little Higgs versus supersymmetry.

\vfill
\noindent --------------- \\
$^\star$ Talk presented  at SUSY 04, Jun 17-23, Tsukuba, Japan\\
 --- submission for the proceedings. 

\clearpage
\addtocounter{page}{-1}

\title{Fermion Sector of  Little Higgs} 

\author{OTTO C.W. KONG }

\address{Department of Physics, National Central University, Chung-li, TAIWAN 32054
\\ {\rm E-mail:  otto@phy.ncu.edu.tw}}

\abstract{The little Higgs mechanism provides an alternative solution to
the hierarchy problem, arguably fitting better into the phenomenological
hint of the "little hierarchy" which may cause some fine-tuning for
the case of supersymmetry. We discuss an aspect of little Higgs physics
lacking proper attention --- the construction of an interesting and consistent 
chiral fermionic sector and its phenomenological implications. At least for 
the kind of example models to be discussed, the gauge and top sector 
structure of a model largely dictates, through gauge anomaly cancellation 
conditions, a specific chiral fermion spectrum. The spectrum has 
interesting, family non-universal, flavor structure. The implications for 
flavor physics are specially interesting.  We also add a brief comment 
of little Higgs versus supersymmetry.
}

\normalsize\baselineskip=15pt

\section{Introduction}
The SM is a model of interactions dictated by an
$SU(3)_C\times SU(2)_L\times U(1)_Y$ gauge symmetry, with a anomaly
free chiral fermion spectrum and a Higgs multiplet responsible for the
spontaneous breaking of the electroweak (EW) symmetry $SU(2)_L\times U(1)_Y$.
Supersymmetry with or without grand unification is the most popular candidate
theory beyond the SM. There,  the beautiful boson-fermion
symmetry to tackle the hierarchy problem, essentially extending the chiral
nature of the fermions to fix the problem for the scalar sector. The approaches 
do not provide any new insight into the difficult problem of the origin of flavor 
structure. Why there are three families of SM fermions is still a fundamental 
problem that we have no credible approach to handle. The so-called
little Higgs mechanism~\cite{acg} comes as an alternative solution to the 
hierarchy problem, with an extended EW symmetry. Here, one can also use
gauge anomaly cancellation constraints to `predict' the fermion spectrum.
The later may even provide an understanding of why three SM 
families, together with specific implications on flavor physics~\cite{mine}.

\section{The Simplest Model}
We focus on the simplest model here. A little Higgs model with a
$SU(3)_L\times U(1)_X$ extended EW symmetry is available~\cite{KS}.
We draw attention to the gauge anomaly considerations and present
the solution spectrum, as given in the table. Note that the embedding of
the SM doublets is not family universal. This is necessary to avoid 
accumulation of $SU(3)_L$ anomalies. Cancellation of the latter is here
achieved by exploring the equality of the number of family and the number
of color, as done in Ref.\cite{331}. The $U(1)$ related gauge anomaly 
contributions are illustrated explicitly. The full fermion spectrum is
essentially fixed by the anomaly cancellation scheme and the little Higgs
requirement of having the extra heavy top quark $T$ living in a $SU(3)$
multiplet with the SM $(t,b)$ doublet.

\begin{table}[t]
\begin{center}\footnotesize
\noindent
 The $SU(3)_C\times SU(3)_L\times U(1)_X$ spectrum with 
little Higgs. Electroweak doublets are put in [.]'s. \\[.05in]
\begin{tabular}{|c|r|r|r|r|r|cc|}
\hline\hline
 & \multicolumn{5}{|c|}{Gauge anomalies} &  \multicolumn{2}{|c|}{$U(1)_Y$ states } \\ 
\hline 
    &  	 $tX$ & $LLL$	& $LLX$ &  $CCX$ &	$X^3$ &				& \\
\hline								  		
${\bf (3_{\scriptscriptstyle C},3_{\scriptscriptstyle L} ,\frac{1}{3})}$
      &    $3$ &	$3$  &	$1$ &	$1$ &  $\frac{1}{3}$	 &  ${\bf \frac{1}{6}}$[$Q$]               & ${\bf \frac{2}{3}}$($T$)    \\
2\ ${\bf ({3}_{\scriptscriptstyle C},\bar{3}_{\scriptscriptstyle L} ,0)}$
     &       0 &	${-6}$ & 	$0$	 & 0 &	0     &  2\ ${\bf \frac{1}{6}}$[$Q$]               & 2\ ${\bf \frac{-1}{3}}$($D,S$)     \\
$3\ {\bf (l_{\scriptscriptstyle C} ,3_{\scriptscriptstyle L} ,\frac{-1}{3})}$
    &	 $- 3$&	$3$  &	   $-1$ &	 & $\frac{-1}{3}$	           & 3\  ${\bf \frac{-1}{2}}$[$L$]          	& 3\ {\bf 0}($N$)   \\
$4\ {\bf (\bar{3}_{\scriptscriptstyle C},1_{\scriptscriptstyle L} ,\frac{-2}{3})}$
   &	$-8$&	 &	&	 $\frac{-8}{3}$ & $\frac{-32}{9}$        & \multicolumn{2}{|c|}{4\ ${\bf \frac{-2}{3}}$ ($\bar{u}, \bar{c}, \bar{t}, \bar{T}$)} 	\\
$5\ {\bf (\bar{3}_{\scriptscriptstyle C},1_{\scriptscriptstyle L} ,\frac{1}{3})}$  
   &	$5$&	&	&	 $\frac{5}{3}$ & $\frac{5}{9}$        & \multicolumn{2}{|c|}{5\ ${\bf \frac{1}{3}}$ ($\bar{d}, \bar{s}, \bar{b}, \bar{D}, \bar{S}$)}\\
$3\ {\bf (1_{\scriptscriptstyle C},1_{\scriptscriptstyle L} ,1)}$   	
    &     $3$  &	 &	&	 &	$3$           &    \multicolumn{2}{|c|}{3\ ${\bf 1}$  ($e^+, \mu^+, \tau^+$)   }       \\ 
 \hline
\multicolumn{1}{|r|}{ Total}    	&	0  &   0  &	0  &	0  & 	0	& 				& \\	
\hline\hline
\end{tabular}
\end{center}
\normalsize
\end{table}

The little Higgs mechanism as a solution to the hierarchy problem only  alleviates
the quadratic divergent quantum correction to the SM Higgs states and admits a natural 
little hierarchy between the EW scale and a higher scale of so-called UV-completion at 
around the 10 TeV order, beyond which further structure would be required. The idea is 
a rather humble bottom-up approach then; but experimental hints
at the existence of such a little hierarchy has been discussed\cite{Bar}.

The little Higgs mechanism is to be implemented here with two scalar multiplets,  
$\Phi_{\!\scriptscriptstyle 1}$ and $\Phi_{\!\scriptscriptstyle 2}$, having the 
right quantum number to couple to the chiral parts of the $T$ quark. This is
illustrated by top-sector Yukawa couplings
\beqa 
{\mathcal L}_{\rm\tiny top}  &=&   \lambda^t_{\!\scriptscriptstyle 1}\,\bar{t'}\,
\Phi_{\!\scriptscriptstyle 1} \, Q   +    \lambda^t_{\!\scriptscriptstyle 2}\,\bar{T'}\,
\Phi_{\!\scriptscriptstyle 2} \, Q
\nonumber \\	&=&
f\,(\lambda^t_{\!\scriptscriptstyle 1}\,\bar{t'}+  \lambda^t_{\!\scriptscriptstyle 2}\,\bar{T'})\, T + \frac{i}{\sqrt{2}} \, (\lambda^t_{\!\scriptscriptstyle 1}\,\bar{t}'
-  \lambda^t_{\!\scriptscriptstyle 2}\,\bar{T}')\,h
\left( \begin{array}{c} t \\ b \end{array} \right) + \cdots
\eeqa
where $Q$ denotes the $(T, t, b)$ triplet (contrary to notation in the table above). 
The SM Higgs doublet $h$ is retrieved as the pseudo-Nambu-Goldstone boson 
from a  $[SU(3)/SU(2)]^2$ nonlinear sigma model parametrization of
$\Phi_{\!\scriptscriptstyle 1}$ and $\Phi_{\!\scriptscriptstyle 2}$. 
The multiplets are required to have aligned $SU(3)$ breaking VEVs, from a
scalar potential with the $[SU(3)]^2$ global symmetry.

\section{General Relevance of the Anomaly Cancellation Considerations}
We are here talking about an extended EW symmetry model as a TeV scale 
effective field theory. This leads to the thinking that may be one needs not 
be asking for cancellation of the gauge anomalies among the chiral fermionic
states. The are some good reasons, however, that make us consider the issue
very relevant.

First, let us take a look at the SM itself, which is in this case an effective
field theory below the scale of the breaking of part of the extended EW
symmetry. The SM has three families of chiral fermionic states each consists
of a prefect unique set of 15 states  with all gauge anomalies well canceled.
Indeed, we have argued in earlier works~\cite{unc67} that one can essentially
derive the spectrum by simply imposing the anomaly cancellation constraints.
This is the best we come to in terms of understanding why there is what there is,
though it still begs the question of why three families
\footnote{Similar considerations have been used to derive minimal chiral spectra 
of bigger gauge symmetries, such as $SU(4)\times SU(3)\times SU(2)\times U(1)$,
with admissible spontaneous symmetry breaking to that of the SM giving
rise to exact the three families as the only remaining chiral states~\cite{unc67}.
}.
It seems very difficult to convince oneself that this is rather some sort of
accident. If the SM fermion spectrum is a guideline, gauge anomaly 
cancellation is relevant. While consistent models with the anomaly or its
cancellation implemented beyond simply the chiral fermionic contributions
are possible, one with an anomaly free fermion spectrum looks far more
attractive Besides, giving up the requirement, one loses an control on what
are the plausible extra fermionic states. Any spectrum may then look as good
as another from a pure theoretical point of view.

From our perspective, little Higgs models of extended EW symmetries can be
constructed to be essentially unique and very predictive so long as the
symmetry is chosen. The gauge symmetry fixes the gauge sector, as well as
the fermionic sector through the anomaly cancellation constraints. Implementation 
of the little Higgs mechanism to stabilize the (little) hierarchy helps to fix the
scalar or Higgs sector. Then, one arrives at a very definite model of TeV scale
physics, without much room for simple modifications, to be checked with 
phenomenological studies. That is a very solid approach for model-building,
with hardly a competing alternative.

\section{A $SU(4)$ Analog}
The basic construction strategy of the anomaly free spectrum may actually be
generalized to the case of any $SU(N)_L\times U(1)_X$ extended EW symmetries.
In the case of $N=4$, it looks like they could be more choice for little Higgs
model-building. A simple extension of the above $SU(3)$ case with one more $T$
quark fits the similarly extended Higgs sector structure. The latter may have a
better behaved Higgs quartic coupling~\cite{KS}. One other fermion spectrum
we find intriguing is presented in the table below. It has a kind of duplicated
fermion list at the QCD/QED level. Each SM fermion get a heavy singlet partner. 
A little Higgs model based on the spectrum would likely have a few special
and phenomenologically appealing features. We have to refrain from elaborating
here though.

\begin{table}[h]
\begin{center} 
{The $SU(4)_L\times U(1)_X$ Spectrum}  \\[.05in]
\begin{tabular}{|c|ccc|}
\hline\hline
				& \multicolumn{3}{|c|}{$U(1)_Y$-states}			\\  \hline
${\bf (3_{\scriptscriptstyle C},4_{\scriptscriptstyle L} ,\frac{1}{6})}$                 &  ${\bf \frac{1}{6}}$[$Q$]               & ${\bf \frac{2}{3}}$($T$)  &      ${\bf \frac{-1}{3}}$($B$) 	\\
2\ ${\bf ({3}_{\scriptscriptstyle C},\bar{4}_{\scriptscriptstyle L} ,\frac{1}{6})}$     &  2\ ${\bf \frac{1}{6}}$[2\ $Q$]               & 2\ ${\bf \frac{-1}{3}}$($D,S$)     & 2\ ${\bf \frac{2}{3}}$($U,C$) \\
$3\ {\bf (l_{\scriptscriptstyle C} ,4_{\scriptscriptstyle L} ,\frac{-1}{2})}$           & 3\  ${\bf \frac{-1}{2}}$[3\ $L$]          	& 3\ {\bf 0}(3\ $N$) 		 & 3\ {\bf -1}(3\ $E^-$)   \\
$6\ {\bf (\bar{3}_{\scriptscriptstyle C},1_{\scriptscriptstyle L} ,\frac{-2}{3})}$        & \multicolumn{2}{|c}{4\ ${\bf \frac{-2}{3}}$ ($\bar{u}, \bar{c}, \bar{t}, \bar{T}$)}     & {2\ ${\bf \frac{-2}{3}}$ ($\bar{U}, \bar{C}$)} 	          	\\
$6\ {\bf (\bar{3}_{\scriptscriptstyle C},1_{\scriptscriptstyle L} ,\frac{1}{3})}$        & \multicolumn{2}{|c}{5\ ${\bf \frac{1}{3}}$ ($\bar{d}, \bar{s}, \bar{b}, \bar{D}, \bar{S}$)} 	& ${\bf \frac{1}{3}}$ ($\bar{B}$)                 	\\
$6\ {\bf (1_{\scriptscriptstyle C},1_{\scriptscriptstyle L} ,1)}$              &    \multicolumn{2}{|c}{3\ ${\bf 1}$  ($e^+, \mu^+, \tau^+$)   }      	 & 3\ {\bf 1}(3\ $E^+$)   	\\
\hline\hline
\end{tabular}
\end{center}
\end{table}

\section{Remarks}
We observe that little Higgs models typically have extended EW symmetries with
extra chiral fermionic states. The gauge anomaly issue comes in. Imposing the
requirement that all gauge anomalies be canceled among the chiral fermionic
states, as in the SM case, has the power to essential dictate the full spectrum. 
Such a complete model has a specific flavor structure which, at least in the case
of $SU(N)_L\times U(1)_X$ symmetries, put the three SM families into one
whole anomaly free set. This may shred new light into the origin of the families.
The full gauge quantum numbers may then be used to extract admissible
Yukawa couplings. As illustrated for the simplest cases of the $SU(3)$ model 
discussed above~\cite{mine},  while the extra singlet quark states can all
obtain consistent TeV scale masses, the generic admissible mass mixings
with the SM quarks pose interesting FCNC constraints. The latter should be
an important aspects of phenomenological explorations of such little Higgs models.

A conceptual comparison of little Higgs versus supersymmetry is in order here.
The former uses (global) bosonic symmetries instead of (local) supersymmetry
to achieve the stability of the SM Higgs mass, hence EW scale. The basic 
quantum field content of the SM is almost perfection, apart from the strange
triplication of the unique anomaly free fermion spectrum. But the scalar/Higgs
field bears the major short-coming --- the hierarchy problem. In the way,
supersymmetry cures the sick scalar sector by pairing it with the healthy (chiral)
sector of fermions. However, the $\mu$-problem makes exactly the conceptual 
incompleteness of the task. The little Higgs scheme uses less specific bosonic
symmetries with only a little accomplishment --- stabilizing only a little hierarchy.
It looks like it has less intuitive beauty. While supersymmetry has nothing to provide
on improving our understanding of the flavor problems, our perspective of
complete little Higgs model construction does better in the aspect. The fermionic
spectrum, though not as beautiful as that of  one SM family, does tied the three
families together in a unique framework.

\section{Acknowledgements}
The author's work is partially supported by the National Science Council of Taiwan
under research grant number NSC 92-2112-M-008-044 and NSC 93-2112-M-008-020.

\bibliographystyle{plain}

\begin{thebibliography}{99}
\bibitem{acg}
N. Arkani-Hamed {\it et.al.}, {\it Phys. Rev. Lett.} {\bf 86}, 4757 (2001).
\bibitem{mine}
O.C.W. Kong,  hep-ph/0307250 (see also  NCU-HEP-k015);
 hep-ph/0308148; hep-ph/0312060, talk given at ICFP II.
\bibitem{KS}
D.E. Kaplan and M. Schmaltz, {\it JHEP} {\bf 0310}, 039 (2003);
M. Schmaltz, hep-ph/0407143.
\bibitem{331}
P.H. Frampton,  {\it Phys. Rev. Lett.} {\bf 69}, 2889 (1992).
\bibitem{Bar}
R. Barbieri, hep-ph/0312253.
\bibitem{unc67}
O.C.W. Kong, {\it Mod. Phys. Lett.} {\bf A11}, 2547 (1996);
{\it Phys. Rev.} {\bf D55}, 383 (1997).


\end{thebibliography}

\end{document}